\documentclass[11pt,a4paper]{article}
\usepackage{jinstpub}
\usepackage{hyperref}

\title{The OctoPAES board for the validation of KM3NeT data acquisition system}
\author[a,b,1]{F.~Filippini%
\note{Corresponding author.}}
\author[b]{T.~Chiarusi,}
\author[b]{G.~Balbi,}
\author[b]{G.~Pellegrini,}
\author[a,b]{F.~Benfenati,}
\author[b]{L.~Degli Esposti,}
\author[c]{D.~Real,}
\author[c]{D.~Calvo,}
\author[d]{V.~Van Beveren}

\affiliation[a]{Dipartimento di Fisica e Astronomia dell’Università di Bologna, Viale Berti Pichat 6/2, 40127 Bologna, Italy}
\affiliation[b]{INFN, Sezione di Bologna, Viale Berti-Pichat 6/2, 40127 Bologna, Italy}
\affiliation[c]{IFIC-Instituto de Física Corpuscular (CSIC - Universitat de València) c/ Catedrático José Beltrán, 2 E-46980 Paterna, Valencia, Spain}
\affiliation[d]{Nikhef, Science Park, Amsterdam, The Netherlands}
\collaboration[c]{on behalf of
KM3NeT collaboration}
\emailAdd{francesco.filippini9@unibo.it}
\abstract{A testbench has been set up at the INFN Sezione di Bologna to optimise key elements of the KM3NeT data acquisition system. A complete framework has been built to simulate a full detection unit and test the optical network, time synchronisation, and on-shore computing resources. A fundamental tool in the test-setup is a customized electronic board: “the OctoPAES”. 
Based on an Altera MAX10 CPLD, it is designed to emulate in a realistic way the optical and acoustic signals recorded by the underwater detectors. This allows us to test, in extreme conditions, the acquisition system and validate its performance with realistic data. If properly configured, the optical data provided by the OctoPAES can be combined to emulate the signals of a through-going muon or other calibration events. In this contribution the OctoPAES boards and some of their use cases at the testbench are presented.}
\keywords{Detector R$\&$D and construction, Data Acquisition System, Neutrino telescope.}
\usepackage{lineno}
\usepackage{lipsum}

\begin{document}
\flushbottom
\maketitle

\section{KM3NeT and data acquisition system}
KM3NeT is a research infrastructure housing the next generation neutrino telescopes, located at the bottom of the Mediterranean Sea. Once completed, it will host a network of Cherenkov detectors, reaching, in its final configuration, an instrumented volume of several cubic kilometres of sea water.
KM3NeT comprises two different instrumented regions, placed
in separate locations. Both detectors will use the same technology and neutrino detection principle, namely a 3D array of photosensors, called \emph{Digital Optical Module (DOM)}, capable of detecting
Cherenkov light produced along the path of relativistic particles emerging from neutrino interactions.
A collection of 18 DOMs connected to an electro-optical cable and arranged along
a vertical structure is called a detection unit.
\subsection{Data acquisition system}
The modular design for building KM3NeT detectors allows for a progressive implementation
and data taking even with an incomplete detector. The same scalable design
is used for the Trigger and Data Acquisition System \cite{TCCP}. In order to reduce the
complexity of the hardware inside the optical modules, the “all data to shore” concept
is adopted. DOMs are therefore submarine nodes
of the global networking infrastructure that, comprising the computing resources of the
shore-station, form a global layer network.
Its main characteristic is its asymmetry, originated after the so-called optical \emph{broadcast} architecture adopted for the global optical infrastructure. It consists
of few downstream optical links broadcasting slow-control commands to the DOMs by
dedicated split points, while upstream links (one per DOM) transport the collected data to the shore station. It aims at best exploiting the number of optical fibres contained in
the many km-long electro-optical cable which connects the shore station with the
detector. Exploiting these connections, a custom White Rabbit Switch fabric is used to achieve the required nanosecond synchronisation among the DOMs.

\section{The Bologna Common Infrastructure}
The Bologna Common Infrastructure (BCI) is a unique setup within the KM3NeT collaboration, able to recreate all the data processing effects due
to a real implementation of the full DAQ chain inside a controlled environment.
It offers the unique opportunity
to test and validate new updates, concerning off-shore electronics, on-shore hardware, network facilities and all software tools assigned to detector control and to data readout. In fact, the nature of KM3NeT detectors, installed at a depth of $\sim$2500 m and $\sim$3500 m below
sea level, and the modularity required for their construction, makes this laboratory a
central node for testing new developments before the final deployment.
The BCI offers a full detection unit size testbench, comprising 18
Central Logic Boards (CLBs) acting as DOMs, 1 CLB acting as Base each with its own power supply electronic board. Real photomultipliers (PMTs) and acoustic sensors are replaced by specific emulation  boards designed for the BCI setup and described in the next section. All the optical connections from
the detection unit to the on-shore station are implemented, and are compliant to the broadcast
scenario. Finally, the BCI hosts the whole network infrastructure with  the same switch fabrics  implemented in a real shore-station.

\section{OctoPAES boards}
\begin{wrapfigure}{r}{5.5cm}
\includegraphics[width=.5\textwidth]{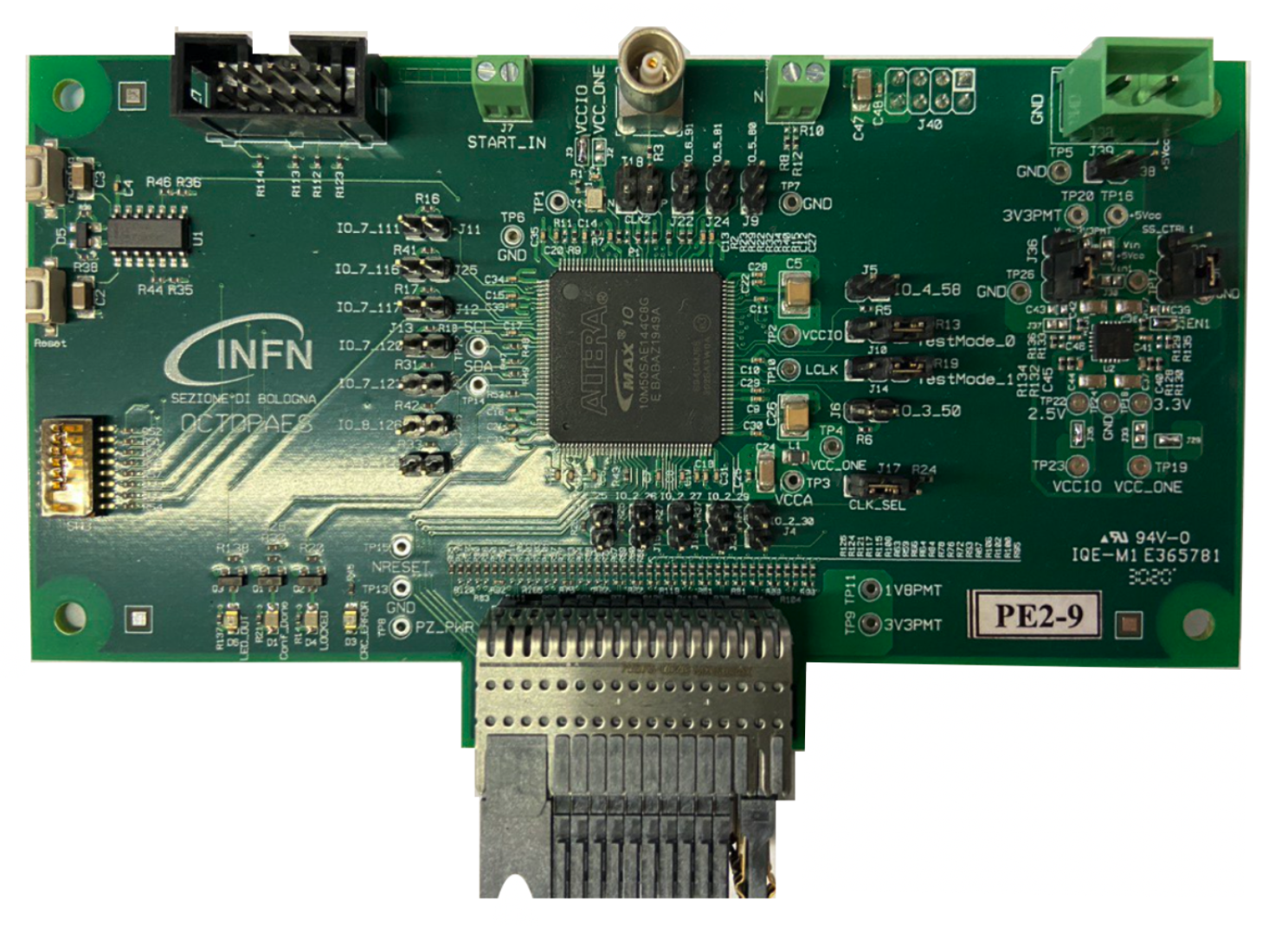}
\caption{In this figure is shown an OctoPAES board, with the specific connector, the DIP switch and all the hardware implementations.}
\label{OP}
\end{wrapfigure}
The OctoPAES boards, shown in Fig.\ref{OP}, were developed by the KM3NeT collaboration together with
the Bologna INFN electronic section, in order to emulate
the behaviour of the corresponding PMTs and Octopus boards of the standard DOM implementation. The name in fact is the acronym for \emph{Photon and Acoustic Emulation of Signals}, outlining therefore the possibility to emulate optical and acoustic signals.
As in the DOM at the BCI testbench each CLB hosts two OctoPAES boards, plugged into the CLB with the same specific connector of the Octopus boards, emulating signals for 19 (OctoPAES LARGE) and 12 (OctoPAES SMALL) PMTs respectively \cite{doi}. The OctoPAES board can be configured in Large or Small mode, thanks to a DIP switch mounted on it. The OctoPAES boards allow to select among different PMT single photoelectron rates and to provide the user with per-PMT configurable hit patterns encoded in the so-called MIF file. In the MIF file,  the sequence of hits emulated per  PMT is represented by a row of 256 digits (i.e. a sequence of 0 and 1).
The OctoPAES transforms each digit into a proper LVDS signal, which is kept above the TDC threshold  when the digit is equal to 1. The readout of the digit sequence is driven by a 80 MHz clock, so each digit equal to 1 corresponds to a hit of 12,5 ns duration. Sequences of digits equal to 1 correspond to hits with a longer duration (multiples of 12,5 ns). A single photoelectron signal is emulated with pairs of ones (for a time over threshold of 25 ns). The MIF file is organised in 128 pages, each containing 32 rows. The first 31 rows correspond to the 31 PMTs of a DOM. The 32th row is left for the implementation of the waveform
for the acoustic emulation (which is not discussed here). The OctoPAES boards can be triggered by an external logic signal and it is also possible to distribute them a common external reference clock. These features allow to start and continuously emulate  hits as originated after the passage of  muons from a precise direction. This is achieved by properly correlating each board MIF file  with  the  appropriate  hit patterns, taking  into account the time and spacial distribution of hits over all the detection unit.  
\subsection{OctoPAES firmware development}
Five different OctoPAES firmware versions have been developed in order to exploit the different functionalities of the board, and to achieve among them time delays less than a maximum of ten nanoseconds:
\begin{itemize}
    \item OctoPAES in Master stand alone mode (i.e. an internal clock is used; the hit emulation
is started and stopped by pressing the appropriate button on the board panel). Allowed single rates per channel:
 5 Hz, 10 Hz, 1 kHz 100 kHz;
\item One Master to many Slave OctoPAES boards. The clock distribution must be implemented as a daisy chain. The start and stop signals are propagated through the clock line;
\item all Slave OctoPAES boards. In this configuration, an external logic board is required to
distribute the clock and the start/stop signal. We chose a Kintex KC705;
\item still all Slave OctoPAES boards and an external logic board as Master, but clock and start/stop
signals distributed through parallel connections;
\item the same hardware setup as in the previous point. In this version of the firmware, the state
machine is modified in order to switch between background and signal hits emulation, properly alternating the appropriate MIF pages. In order to reproduce a standard
condition as measured with the KM3NeT/ARCA telescope, the background hits
are emulated at the rate of 5 kHz while the signal hits are added at 1 or 10 Hz.
\end{itemize}
\section{Muon Emulation}
\begin{figure}[!h]
\centering
\includegraphics[width=0.9\textwidth]{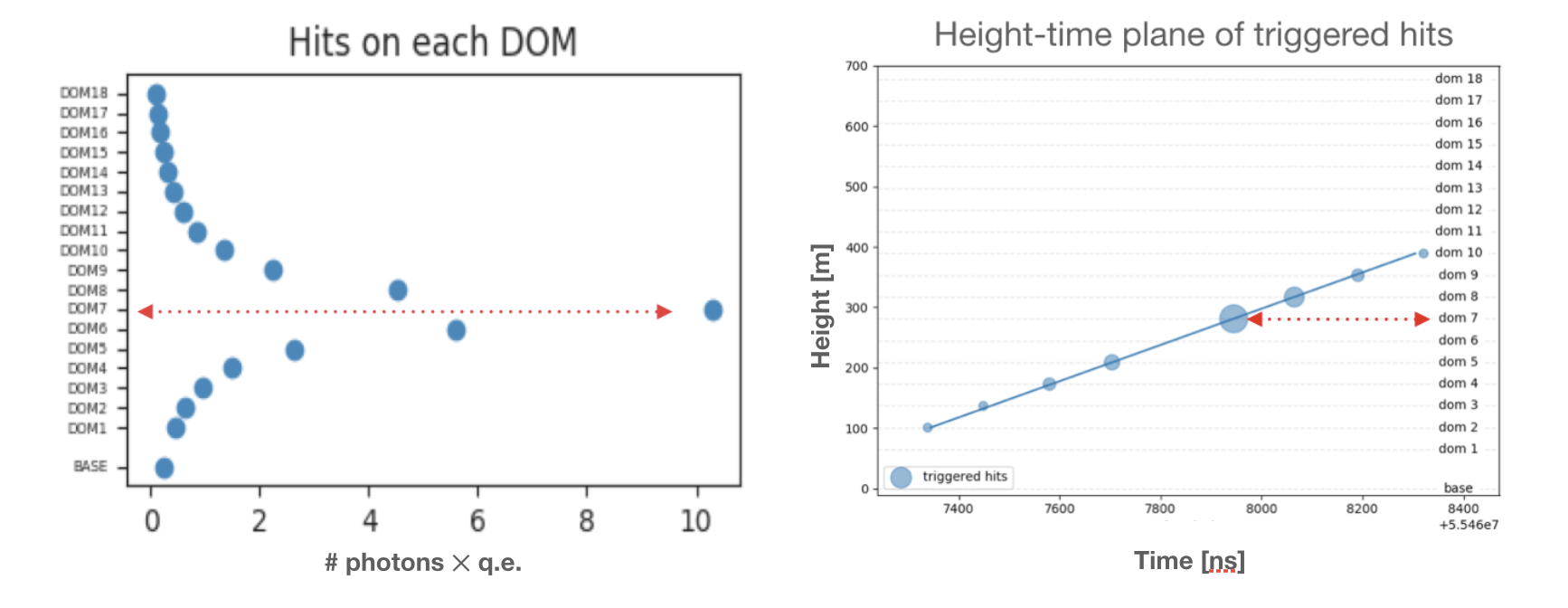}
\caption{The two figures show the signature of the emulated muon. \emph{Left:} calculated distribution of \textit{photons} $\times$ \textit{quantum efficiency} (then rounded to the nearest integer) on single DOMs within one detection unit, used to construct the MIF file; \emph{Right:} arrival time of photon hits, registered in the testbench, in function of the detection unit height. The dots represent the photon hits triggered as event by the DAQ chain. Comparing the two figures, all and only the signal hits ejected by the OctoPAES triggers an event.}
\label{distrooo}
\end{figure}
Since more hit patterns can be concatenated in the MIF
file, the correct signal page can be set via a 12-bit address, selectable by means of a DIP switch device mounted on the OctoPAES board. The above mentioned frequency values,
for both background and signal, are chosen to approximate the ones measured in the
KM3NeT/ARCA deployment site. The consequent background page in the MIF file has been created inserting a 25 ns hit (two consecutive bits=1) in a random location over the 256 bits
in a row. The probability distribution of a hit over the 256 bits is assumed to be uniform. Another useful MIF page type was the \emph{calibration
page}, in which only two coincident hits are inserted on channel 0 (OctoPAES Small) and
on channel 12 (OctoPAES Large). By measuring the time
difference between the available CLBs, the delays of one OctoPAES board with respect to the others can be estimated. Finally, the signal page of each DOM is designed, taking into account  the delays between the photon hits, computed after the predefined trajectory of the incoming muon. The number of hits per DOM and the distribution over the various channels were determined taking into account the distance between the muon trajectory and the detector, the absorption of photons in seawater and a 30$\%$ PMT quantum efficiency (assumed constant over the  optical wavelength range). No scattering effects are considered. Given the muon trajectory, the Cherenkov wavefront was convolved with the spherical distribution of PMTs within the DOM to compute the geometrical acceptance factor. Effect due to different wavelengths were treated according to the Frank-Tamm formula. Once the number of hits on each DOM is calculated, as shown on the left of Fig.\ref{distrooo}, the right PMTs were selected in order to provide the directional information gathered in the multi-PMT configuration.
\section{Conclusion}
The OctoPAES boards extend the possibilities and capabilities of the BCI testbench, offering the possibility to manipulate the payload on the system and to emulate the passage of particles as in a real detector, in order to recreate hit rates fully complaint to the KM3NeT/ARCA environment. The possibilities opened by these boards will allows us, in the future, to test with more accuracy new features and new developments under "almost" real data-taking conditions, but in a controlled and safe environment as the BCI testbench.
\bibliography{lit}
\medskip
\end{document}